\documentclass[12pt]{iopart}
\usepackage{epsfig}

\begin{document}

\title{Cosmological dynamics of scalar fields with O(\emph{N}) symmetry}

\author{Jian-gang Hao and Xin-zhou Li}

\address{Shanghai United Center for Astrophysics(SUCA), Shanghai Normal University, Shanghai 200234, China}
\address{E-mail: kychz@shnu.edu.cn}

\begin{abstract}
In this paper, we study the cosmological dynamics of scalar fields
with O(\emph{N}) symmetry in general potentials. We compare the
phase space of the dynamical systems of the quintessence and
phantom and give the conditions for the existence of various
attractors as well as their cosmological implications. We also
show that the existence of tracking attractor in O(\textit{N})
phantom models require the potential with $\Gamma<1/2$, which
makes the model with exponential potential possess no tracking
attractor.

\end{abstract}

\pacs{0440-b, 9880Cq, 9880Es}



\section{Introduction}

Astronomical measurements from Supernovae \cite{riess, perlmutter,
tonry} and CMB anisotropy \cite{bennett,Netterfield,Halverson}
independently confirm that about two-thirds of the energy density
in our Universe is dark energy, which has negative pressure and
accelerates the expansion of the Universe. The most widely studied
models for dark energy are cosmological constant and a time
varying scalar field with positive or negative kinetic energy
evolving in a specific potential, referred to as "quintessence"
with $w>-1$ \cite{ratra,Wetterich,Ferreira,Coble,
Steinhardt,Peebles} and "phantom" with $w<-1$
\cite{caldwell,schulz,Carroll,hao1,Sami,Odintsov} respectively. Since
current observational constraint on the equation of state of the
dark energy lies in the range $-1.38<w<-0.82$ \cite{Melchiorri},
it is still too early to rule out any of the above candidates.
Complex scalar field as quintessence has been proposed in
\cite{gu, Boyle} and its generalization to the scalar fields with
O(\emph{N}) symmetry as quintessence and phantom in exponential
potential have been done in \cite{li, hao2}. To study the global
property of the cosmological system containing dark energy, phase
space analysis is proved to be a powerful tool. For real scalar
field quintessence and phantom models in FRW Universe, the phase
space analysis have been done in \cite{copeland,wang,Ng,hao3}
while the analysis has been extended to the O(\textit{N}) scalar
fields in exponential potentials in Ref.\cite{li,hao2}. In
exponential potential, it has been show that the O(\textit{N})
phantom does not admit tracking attractor that plays an important
role in alleviating the fine-tuning problem. In this paper, we
study the phase space structure of the scalar fields with
O(\emph{N}) symmetry in general potentials and discuss the various
attractors as well as the conditions for their stability.

\section{Phase space of quintessence with O(\emph{N}) symmetry in general potentials}

In this section, we study the phase space of quintessence with
O(\emph{N}) symmetry in flat FRW cosmological background

\begin{equation}\label{metric}
ds^2=dt^2-a^2(t)d\textbf{x}^2
\end{equation}
The corresponding equations of motion as well as Einstein
equations are just a simple generalization of those for
exponential potential as shown in \cite{li}.
\begin{eqnarray}\label{sys1}
\dot{H}=-\frac{\kappa^2}{2}(\rho_\gamma+p_\gamma+\dot{R}^2 +
\frac{\Sigma^2}{a^6R^2})\\\nonumber
\dot{\rho_\gamma}=-3H(\rho_\gamma+p_\gamma)\\\nonumber
\ddot{R}+3H\dot{R}-\frac{\Sigma^2}{a^6R^3}+V'(R)=0\\\nonumber
H^2=\frac{\kappa^2}{3}[\rho_\gamma+\frac{1}{2}(\dot{R}^2+\frac{\Sigma^2}{a^6R^2})+V(R)]
\end{eqnarray}

\noindent where $R$ is the radial component of the scalar fields
with O(\textit{N}) symmetry and all "angular" equations have been
integrated out and their contributions to the system is reflected
by the terms associated with $\Sigma$, the integration constant of
"angular" equations \cite{li}. $\rho_\gamma$ and $p_\gamma$ are
the energy density and pressure of the baryotropic matter and
$p_\gamma=(\gamma-1)\rho_\gamma$, where $\gamma$ is a constant,
$0\leq \gamma\leq2$. Here, we denote $\kappa^2=8\pi G$. Now,
introducing the following variables:
$x=\frac{\kappa\dot{R}}{\sqrt{6}H}$,
$y=\frac{\kappa\sqrt{V(R)}}{\sqrt{3}H}$,
$z=\frac{\kappa}{\sqrt{6}H}\frac{\Sigma}{a^3R}$,
$\xi=\frac{1}{\kappa R}$, $\lambda=-\frac{V'(R)}{\kappa V(R)}$,
$\Gamma=\frac{V(R)V''(R)}{V'^2(R)}$ and $N=\log a$, the equation
system(\ref{sys1}) becomes the following autonomous system:

\begin{eqnarray}\label{auto1}
\frac{dx}{dN}&=&\frac{3}{2}x[\gamma(1-x^2-y^2-z^2)+2(x^2+z^2)]-(3x-\sqrt{6}z^2\xi-\sqrt{\frac{3}{2}}\lambda
y^2)\\\nonumber
\frac{dy}{dN}&=&\frac{3}{2}y[\gamma(1-x^2-y^2-z^2)+2(x^2+z^2)]-\sqrt{\frac{3}{2}}\lambda
xy\\\nonumber
\frac{dz}{dN}&=&-3z+\frac{3}{2}z[\gamma(1-x^2-y^2-z^2)+2(x^2+z^2)]-\sqrt{6}xz\xi\\\nonumber
\frac{d\xi}{dN}&=&-\sqrt{6}\xi^2x\\\nonumber \frac{d\lambda}{dN}
&=& -\sqrt{6}\lambda^2x(\Gamma-1)
\end{eqnarray}

\noindent Now, it is straightforward to analyze the critical
points as well as their stability. In \textbf{table 1}, we list
the stable critical points, conditions for their existence and the
cosmological parameters there (Note that we have omitted those
unstable critical points). We include a detailed analysis for the
tracking attractor and the corresponding eigenvalues of the linear
perturbation matrix in \textbf{Appendix A} and \textbf{B}.

\clearpage
\begin{center}
\begin{table}
\begin{tabular}{ c c c c c }
  \hline
  Critical points
  (x,y,z,$\xi$) & Name & Conditions  & $\Omega_{R}$ & $w_{R}$ \\
\hline
   $\frac{\lambda}{\sqrt{6}}$,$\sqrt{1-\frac{\lambda^2}{6}}$, 0, 0 & quintessence attractor & $\Gamma=1$, $\lambda\neq 0$, $\lambda^2<3\gamma$ & 1 & $-1+\frac{\lambda^2}{3}$\\
  $\sqrt{\frac{3}{2}}\frac{\gamma_e}{\lambda}$,$\sqrt{\frac{3\gamma_e(2-\gamma_e)}{2\lambda^2}}$,0,0 & tracking attractor & $\Gamma>\frac{1}{2}$,$\lambda\neq 0$, $\gamma_e<2$ & $\frac{3\gamma_e}{\lambda^2}$ & $\gamma_e-1$ \\
  0,1,0,0 & de Sitter attractor & $\lambda=0$ & 1 & -1 \\  \hline

\end{tabular}
\caption{The critical points and their physical properties for
O(\textit{N}) quintessence model. Here, $\gamma_e\equiv
\frac{\gamma}{2\Gamma-1}$. Considering the requirements
$\gamma_e<2$ and $\Gamma>\frac{1}{2}$ together, we have
$\Gamma>\frac{\gamma/2+1}{2}$ for the stability of the attractor.}
\end{table}
\end{center}

\section{Phase space of phantom with O(\emph{N}) symmetry in general potentials}
Phantom with O(\emph{N}) symmetry in exponential potential has
been proposed in Ref.\cite{hao2}. However, in exponential
potential, the tracking attractor does not exist and thus make the
model need careful fine-tuning. In this section, we will consider
the model in a general potential and give a condition for the
existence of tracking attractor. In flat FRW spacetime, the
equations of motion as well as Einstein equations are

\begin{eqnarray}\label{psys1}
\dot{H}=-\frac{\kappa^2}{2}(\rho_\gamma+p_\gamma-\dot{R}^2 -
\frac{\Sigma^2}{a^6R^2})\\\nonumber
\dot{\rho_\gamma}=-3H(\rho_\gamma+p_\gamma)\\\nonumber
\ddot{R}+3H\dot{R}-\frac{\Sigma^2}{a^6R^3}-V'(R)=0\\\nonumber
H^2=\frac{\kappa^2}{3}[\rho_\gamma-\frac{1}{2}(\dot{R}^2+\frac{\Sigma^2}{a^6R^2})+V(R)]
\end{eqnarray}

\noindent By using the same dimensionless parameters as defined in
previous section, we can reduce the equation system (\ref{psys1})
to

\begin{eqnarray}\label{pauto1}
\frac{dx}{dN}&=&\frac{3}{2}x[\gamma(1+x^2-y^2+z^2)-2(x^2+z^2)]-(3x-\sqrt{6}z^2\xi+\sqrt{\frac{3}{2}}\lambda
y^2)\\\nonumber
\frac{dy}{dN}&=&\frac{3}{2}y[\gamma(1+x^2-y^2+z^2)-2(x^2+z^2)]-\sqrt{\frac{3}{2}}\lambda
xy\\\nonumber
\frac{dz}{dN}&=&-3z+\frac{3}{2}z[\gamma(1+x^2-y^2+z^2)-2(x^2+z^2)]-\sqrt{6}xz\xi\\\nonumber
\frac{d\xi}{dN}&=&-\sqrt{6}\xi^2x\\\nonumber \frac{d\lambda}{dN}
&=& -\sqrt{6}\lambda^2x(\Gamma-1)
\end{eqnarray}

\noindent By comparing the dynamical system of O(\textit{N})
phantom Eqs.(\ref{psys1}) with O(\textit{N}) quintessence
Eqs.(\ref{sys1}), one can find that their differences are merely
the sign difference of the terms associated with kinetic energy.
However, these small changes make their dynamics quite different.
In \textbf{table 2}, we list the critical points, conditions of
existence and stability as well as cosmological parameters there
for O(\textit{N}) phantom in general potentials. We need to
emphasize that in exponential potential, the O(\textit{N}) phantom
model does not admit tracking attractor solution\cite{hao2}. This
is due the fact that the existence of tracking attractor requires
that the potentials must have $\Gamma<1/2$ while the exponential
potential with $\Gamma=1$ is excluded. Another important point of
O(\textit{N}) phantom lies in that the existence of stable big rip
attractor requires $\lambda^2<3\gamma$, which therefore impose a
lower bound to the equation of state $w>-1-\gamma$. The similar
property $w<-1+\gamma$ also hold for the O(\textit{N})
quintessence case but the significance is less because this upper
bound won't be much physical interest.

From the above qualitative analysis, we need to stress that the
critical point structure is unchanged from the single-field when
the internal O(\emph{N}) symmetry is introduced since the critical
points correspond to vanishing angular components ($z=0$).
However, the angular component will definitely affect the
trajectories approaching the critical points which we will explore
in the ensuing section.

\begin{center}
\begin{table}
\begin{tabular}{ c c c c c }
  \hline
  Critical points
  (x,y,z,$\xi$) & Name & Conditions  & $\Omega_{R}$ & $w_{R}$ \\
\hline
   $-\frac{\lambda}{\sqrt{6}}$,$\sqrt{1+\frac{\lambda^2}{6}}$, 0, 0 & big rip attractor & $\Gamma=1$, $\lambda\neq 0$, $\lambda^2<3\gamma$ & 1 & $-1-\frac{\lambda^2}{3}$\\
  $\sqrt{\frac{3}{2}}\frac{\gamma_e}{\lambda}$,$\sqrt{\frac{3\gamma_e(\gamma_e-2)}{2\lambda^2}}$,0,0 & tracking attractor & $\Gamma<\frac{1}{2}$,$\lambda\neq 0$, $\gamma_e<2$ & $-\frac{3\gamma_e}{\lambda^2}$ & $\gamma_e-1$ \\
  0,1,0,0 & de Sitter attractor & $\lambda=0$ & 1 & -1 \\
   \hline
\end{tabular}
\caption{The critical points and their physical properties for
O(\textit{N}) phantom model. Here, $\gamma_e\equiv
\frac{\gamma}{2\Gamma-1}$}
\end{table}
\end{center}

\section{Numerical analysis of the O(\textit{N}) quintessence and phantom}

In previous sections, we have studied the phase space of scalar
field with O(\textit{N}) symmetry in a flat FRW cosmological
background. The critical points indicate that these stable
attractor phases corresponding to the vanishing of "angular"
components resulted from the O(\textit{N}) symmetry. Thus, what
will be the effects of the O(\textit{N}) symmetry? In fact, the
introduction of the "angular" term will change the evolutionary
tracks of the dynamical systems, which was not well manifested by
the above qualitative analysis. In this section, we study their
dynamical evolution numerically and compare the results with the
cases of real scalar field models. Here, we choose the potential
as the power law of the field $V(R)=V_0(R_0/R)^q$. We choose $q=4$
and $q=-5/4$ for O(\textit{N}) quintessence and phantom with
corresponding $\Gamma=5/4$ and $\Gamma=1/5$. The parameters were
chosen as $V_0=0.00001$, $R_0=H_0=\kappa=1$ and $\Sigma=15$.

From Fig.1 to Fig.4, we plot the dynamical evolution of matter,
radiation and dark energy be it quintessence or phantom with or
without O(\textit{N}) symmetry. Comparing Fig.1 and Fig.2 as well
as Fig.3 and Fig.4, one can find that the internal symmetry will
affect the dynamical evolution at very early epoch. The early
property of quintessence or phantom will definitely affect the CMB
spectrum and thus provide the possibility that how to discriminate
them from their counterpart without O(\textit{N}) symmetry.
Especially, for O(\textit{N}) quintessence, the energy density is
not necessarily negligible compared to radiation and matter due to
the contribution of "angular" term that is proportional to
$a^{-6}$, which allows  a non-negligible energy density for dark
energy at early epoch. This is a very interesting feature because
if the dark energy is cosmological constant, its initial energy
density need to be some 100 orders of magnitude or smaller than
the initial matter-energy density \cite{muk} and thus pose the
fine-tuning problem and coincidence problem. While in
O(\textit{N}) quintessence, the "angular" term will contribute a
kinetic energy that decrease very fast as the universe expands
(proportional to $a^{-6}$). So, the initial energy condition could
be comparable to the matter which is more natural than the real
scalar field models.

\begin{figure}[b]
\begin{center}
\epsfig{file=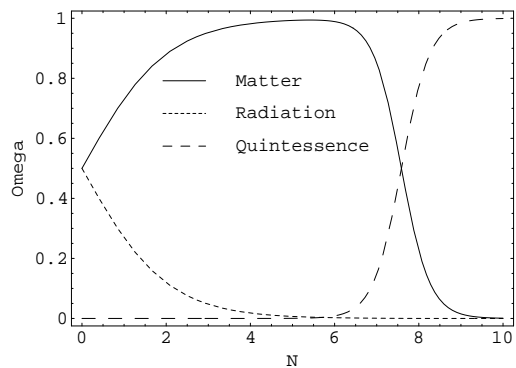,height=2in,width=2.8in} \caption{Evolution
of cosmic parameters for matter, radiation and quintessence
without O(\textit{N}) symmetry. The plot begins with
$\Omega_{m,i}=\Omega_{r,i}=0.499$ and $\Omega_R=0.002$. Note also
that the label \emph{N} of horizontal axis has nothing to do with
the O(\emph{N}) symmetry, it is the e-fold number defined as
$N=\ln a$. }
\end{center}
\end{figure}

\begin{figure}[b]
\begin{center}
\epsfig{file=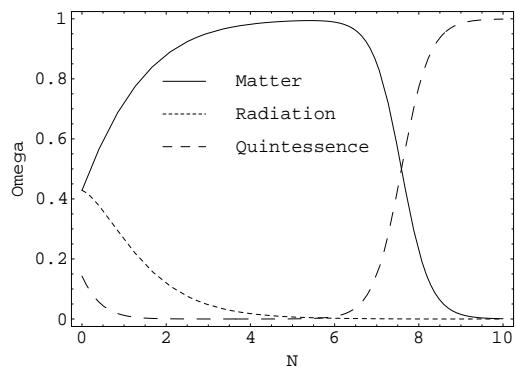,height=2in,width=2.8in} \caption{Evolution
of cosmic parameters for matter, radiation and quintessence with
O(\emph{N}) symmetry. The tilt of the curve at the early epoch is
an important feature of O(\textit{N}) quintessence. The plot
begins with $\Omega_{m,i}=\Omega_{r,i}=0.425$ and
$\Omega_R=0.150$}
\end{center}
\end{figure}

\clearpage
\begin{figure}
\begin{center}
\epsfig{file=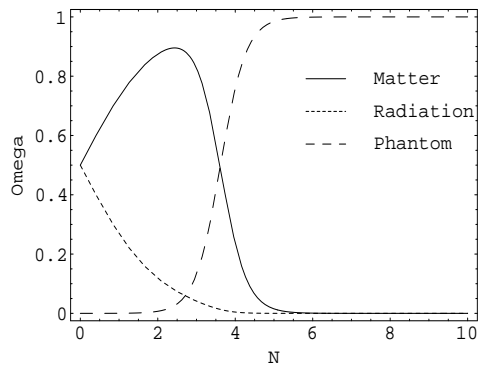,height=2in,width=2.8in} \caption{Evolution
of cosmic parameters for matter, radiation and phantom without
O(\textit{N}) symmetry. The plot begins with
$\Omega_{m,i}=\Omega_{r,i}=0.499$ and $\Omega_R=0.002$}
\end{center}
\end{figure}

\begin{figure}
\begin{center}
\epsfig{file=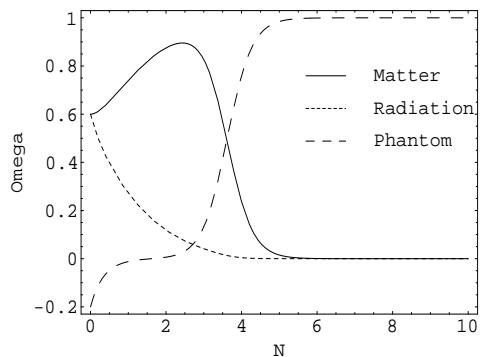,height=2in,width=2.8in} \caption{Evolution
of cosmic parameters for matter, radiation and phantom with
O(\emph{N}) symmetry. To make the difference between the
O(\textit{N}) phantom and phantom clearer, we alow a negative
$\Omega_{R}$ at the initial part, which by definition is positive.
In fact, in a realistic model, the initial values should be chosen
with respect to the energy condition $\rho\geq 0$ and the result
will be that all curves will go upwards. }
\end{center}
\end{figure}

\section{Discussion and Conclusion }

In this paper, we study the cosmological dynamics of scalar field
with O(\textit{N}) symmetry in general potentials. We show that
although there are merely several sign differences between the
O(\textit{N}) phantom and quintessence system, the dynamical
evolution, the phase space property and the physical significance
are quite different, which lead to the difference requirements for
the potentials to admit stable tracking attractors. This
requirement for O(\textit{N}) phantom, i.e. $\Gamma<1/2$ makes the
exponential potential with $\Gamma=1$ admit no tracking attractor,
which is in agreement with the conclusion obtained in
Ref.\cite{hao2}. Another important point we need to stress is that
in the phantom model, the stability of big rip attractor requires
that the resulting equation of state $w>-1-\gamma$, which has very
interesting physical implication because this may add a natural
priori for the lower bound of $w$ and plays a very important role
in data analysis of SNeIa\cite{tonry}. We also show that for the
O(\emph{N}) quintessence model, the initial energy density could
be larger and won't affect the evolution too much because the
"angular" kinetic energy decreases with $a^{-6}$. In real scalar
field quintessence, tracker mechanism was used to provide a
non-negligible energy density at early epoch. In the O(\textit{N})
quintessence model, if equipped with the same tracker mechanism,
it will admit a wider range of initial energy density.

\appendix
\section{Analysis for Tracking attractors of O(\textit{N}) quintessence and phantom }

In this appendix, we show the way to analyze the tracking
attractors in the dynamical systems of O(\textit{N}) quintessence
and phantom. From the dynamical system listed in \textbf{section}
\textbf{2} and \textbf{3}, one can not identify tracking
attractors. The tracking attractors listed in \textbf{Table 1} and
\textbf{Table 2} are obtained through the following analysis. The
tracking attractors exist when $\lambda$ is very large and
$\Gamma\simeq constant$. So, we introduce the following
dimensionless variables re-scaled by
$\epsilon\equiv\frac{1}{\lambda}$. These variables are
$X=x/\epsilon$, $Y=y/\epsilon$, $Z=z/\epsilon$ and
$\Theta=\xi/\epsilon$ with $x$, $y$, $z$ and $\xi$ defined in
\textbf{section 2}. Then the dynamical system could be rewritten
as

\begin{eqnarray}\label{auto2}
\frac{dX}{dN}&=&-\sqrt{6}X^2(\Gamma-1)+\frac{3}{2}\gamma X-3
X\pm\sqrt{\frac{3}{2}}Y^2\\\nonumber
\frac{dY}{dN}&=&-\sqrt{6}XY(\Gamma-1)+\frac{3}{2}\gamma
Y-\sqrt{\frac{3}{2}}XY\\\nonumber
\frac{dZ}{dN}&=&-\sqrt{6}XY(\Gamma-1)+\frac{3}{2}\gamma
Z-3Z\\\nonumber \frac{d\Theta}{dN}&=&-\sqrt{6}X\Theta(\Gamma-1)
\end{eqnarray}

\noindent where the plus and minus signs of the last term of the
first equation correspond to quintessence and phantom
respectively. Also, we have dropped the terms containing the small
quantity $\epsilon$ in Eqs.(\ref{auto2}).  Now, it will be
straightforward to obtain the critical points of the above
equation system (\ref{auto2}) by setting the righthand sides to
zero. The results are contained in \textbf{table 1} and
\textbf{table 2}. It is worth noting that the equations for
O(\textit{N}) quintessence and for O(\textit{N}) phantom are very
similar with only a sign difference of one term. However, this
small change lead to very different expressions for the cosmic
density parameters at the tracking attractors, i.e.
$\frac{3\gamma_e}{\lambda^2}$ and $-\frac{3\gamma_e}{\lambda^2}$
for quintessence and phantom respectively. These expressions
indicate that $\gamma_e$ for O(\textit{N}) phantom must be
negative so that the corresponding $\Omega_R$ makes sense and thus
requires that $\Gamma$ of the potential must be less than $1/2$.
From the definition of $\Gamma$, it would be clear that
exponential potential has $\Gamma=1$ and therefore won't admit
tracking attractor in phantom dynamical system.

\clearpage

\section{The eigenvalues of the linear perturbation matrix for different stable critical points}

To determine the stability, we need to linearize the corresponding
equation system (\ref{auto1}), (\ref{pauto1}) and (\ref{auto2}) in
a neighborhood of the critical points, and then determine the
stability of the systems by the eigenvalues of the coefficients
matrix. By this way, we calculated the various eigenvalues
corresponding to each critical points and tabulated the results in
\textbf{table B1} and \textbf{table B2}. We need to point out that
the eigenvalules for tracking attractors of O(\textit{N})
quintessence and phantom are the same to each other because the
linearized equations have only one sign-different term, see
Eqs.(\ref{auto2}) and this difference just does not enter into the
corresponding eigenvalues coincidentally. From the eigenvalues, we
obtain the condition for the stability of the critical points,
which is contained in \textbf{Table 1} and \textbf{Table 2}. Note
that the eigenvalues for the tracking attractors are very
complicated, and thus we compute them for the power law potential
as we have used in \textbf{section 4}. The results are listed in
\textbf{Table B3}, from which we know that the tracking attractors
are stable.

\begin{table}[b]
\begin{tabular}{ c c }
  \hline
 Critical points (x, y, z, $\xi$) & eigenvalues \\
  \hline
  $\frac{\lambda}{\sqrt{6}}$,$\sqrt{1-\frac{\lambda^2}{6}}$, 0, 0  &  0, $\frac{1}{2}(\lambda^2-6)$, $\frac{1}{2}(\lambda^2-6)$, $\lambda^2-3\gamma$\\
   & \\
  $\sqrt{\frac{3}{2}} \gamma_e$, $\sqrt{\frac{3\gamma_e(2-\gamma_e)}{2}}$,0,0  & \small{
  $-3(\Gamma-1)\gamma_e$, $\frac{-3(2+(2\Gamma-3)\gamma_e)}{4}[1+\frac{\sqrt{4+(4-24\Gamma)\gamma_e+(1+2\Gamma)^2\gamma_e^2}}{2+(2\Gamma-3)\gamma_e}]$} \\
     & \small{$\frac{3}{2}(\gamma_e-2)$, $\frac{-3(2+(2\Gamma-3)\gamma_e)}{4}[1-\frac{\sqrt{4+(4-24\Gamma)\gamma_e+(1+2\Gamma)^2\gamma_e^2}}{2+(2\Gamma-3)\gamma_e}]$}\\
      & \\0, 1, 0, 0 & -3, -3, 0, $-3\gamma$\\
  \hline
\end{tabular}
\caption{The eigenvalues of the critical points for O(\textit{N})
quintessence. Note that the critical point for the tracking
attractor (the second row) as well as its corresponding
eigenvalues are given in terms of (X, Y, Z, $\Theta$)}
\end{table}

\begin{table}[b]
\begin{tabular}{ c c }
  \hline
 Critical points (x, y, z, $\xi$) & eigenvalues \\
  \hline
 $\frac{-\lambda}{\sqrt{6}}$,$\sqrt{1+\frac{\lambda^2}{6}}$, 0, 0&   0, $\frac{1}{2}(-\lambda^2-6)$, $\lambda^2-3\gamma$, $\frac{1}{2}(\lambda^2-6)$ \\
   & \\
   $\sqrt{\frac{3}{2}}\gamma_e$, $\sqrt{\frac{3\gamma_e(\gamma_e-2)}{2}}$, 0, 0  & \small{
  $-3(\Gamma-1)\gamma_e$, $\frac{-3(2+(2\Gamma-3)\gamma_e)}{4}[1+\frac{\sqrt{4+(4-24\Gamma)\gamma_e+(1+2\Gamma)^2\gamma_e^2}}{2+(2\Gamma-3)\gamma_e}]$} \\
   & \small{$\frac{3}{2}(\gamma_e-2)$, $\frac{-3(2+(2\Gamma-3)\gamma_e)}{4}[1-\frac{\sqrt{4+(4-24\Gamma)\gamma_e+(1+2\Gamma)^2\gamma_e^2}}{2+(2\Gamma-3)\gamma_e}]$}\\
   & \\
   0, 1, 0, 0 & -3, -3, 0, $-3\gamma$ \\
  \hline
\end{tabular}
\caption{The eigenvalues of the critical points for O(\textit{N})
phantom. Note that, similar to the case in Table B1, the critical
point for the tracking attractor (the second row) as well as its
corresponding eigenvalues are given in terms of (X, Y, Z,
$\Theta$)}
\end{table}

\begin{table}[t]
\begin{center}
\begin{tabular}{c c c c c c}
  \hline
Models  &$\Gamma$ & $\gamma$ & $\gamma_e$ & $ Re(\lambda_1)$ & $Re(\lambda_2)$ \\
  \hline
O(\textit{N}) quintessence &5/4 & 4/3 & 8/9 & -1.1667 & -1.1667 \\
 &  & 1 & 2/3 & -1.2500 & -1.2500 \\\hline
 O(\textit{N}) phantom & 1/5 & 4/3 & -20/9 & -8.7820 & -2.8847 \\
  &  & 1 & -5/3 & -7.2122 & -2.2878 \\
  \hline
\end{tabular}
\caption{The numerical results of the real part of the eigenvalues
for the tracking attractors when the background fluid is radiation
$\gamma=4/3$ and matter $\gamma=1$ in the power law potential as
we described in \textbf{section 4}.}
\end{center}
\end{table}

\ack{This work is supported by NKBRSF under Grant No. 1999075406.}


\section*{References}
\begin{thebibliography}{99}
\bibitem{riess} Riess A G
\textit{et al.} 1998 \emph{Astron. J}. {\bf 116} 1009
\bibitem{perlmutter} Perlmutter S
\textit{et al.} 1999 \emph{Astrophys. J}. {\bf 517} 565
\bibitem{tonry} Tonry J L
\textit{et al.} 2003 \emph{Astrophys. J}. \textbf{594} 1
\bibitem {bennett} Bennett C L \textit{et al.} \emph{Astrophys.
J}. Suppl. \textbf{148} 1
\bibitem{Netterfield} Netterfield C B \textit{et al.} 2002 \emph{Astrophys.
J}. \textbf{517} 604
\bibitem{Halverson} Halverson N W \textit{et al.} 2002
\emph{Astrophys. J}. \textbf{568} 38
\bibitem {ratra} Ratra B and Peebles P J
1988 \emph{Phys. Rev}. {\bf D37} 3406
\bibitem {Wetterich}Wetterich C 1988 \emph{Nucl.
Phys.} \textbf{B302} 668
\bibitem {Ferreira}Ferreira P and Joyce M 1997 \emph{Phys. Rev. Lett}. \textbf{79}
4740
\bibitem {Coble} Coble K, Dodelson S and
Frieman J 1997 \emph{Phys. Rev}. {\bf D55} 1851
\bibitem {Steinhardt} Caldwell R R, Dave R
 and Steinhardt P J 1998 \emph{Phys. Rev. Lett}. {\bf 80} 1582
\bibitem {Peebles}
Peebles P J E, Ratra B 2003 \emph{Rev. Mod. Phys}. \textbf{75} 599
\\ Padmanabhan T 2003 \emph{Phys. Rep}. \textbf{380} 235
\bibitem {caldwell} Caldwell R R 2002 \emph{Phys. Lett}. \textbf{B545} 23
\bibitem {schulz} Schulz A E and White M 2001 \emph{Phys. Rev}. \textbf{D64}
043514
\bibitem {Carroll} Carroll S M, Hoffman M and Trodden M 2003 \emph{Phys.Rev}.\textbf{D68}
023509
\bibitem {hao1}
Hao J G and Li X Z 2003 \emph{Phys. Rev.} \textbf{D67} 107303
\\ Hao J G and Li X Z 2003 \emph{Phys. Rev}. \textbf{ D68} 043501
\bibitem {Sami} Singh P,
Sami M and Dadhich N 2003 \emph{Phys. Rev.} \textbf{D68} 023522
\bibitem {Odintsov} Nojiri S and Odintsov S D 2003 \emph{Phys. lett.} \textbf{B562} 147
\\ Elizalde E, Nojiri S and Odintsov S D \emph{hep-th/0405034}; \emph{hep-th/0408170}
\bibitem {Melchiorri} Melchiorri A, Mersini L, Odman C J and
Trodden M 2003 \emph{Phys. Rev}. \textbf{D68} 043509
\bibitem {gu} Gu J A and Huang W Y 2001 \emph{Phys.Lett}. \textbf{B517} 1
\bibitem {Boyle}
Boyle L A, Caldwell R R and Kamionkowski M 2002 \emph{Phys. Lett}.
\textbf{B545} 17
\bibitem {li}Li X Z, Hao J G and Liu D J 2002 \emph{Class.Quant.Grav.} \textbf{19} 6049
\bibitem{hao2}Li X Z and Hao J G 2004 \emph{Phys. Rev}.
\textbf{D69} 107303

\bibitem{copeland} Copeland E J, Liddle A R and
Wands D 1998 \emph{Phys.Rev}. \textbf{D57} 4686
\bibitem{wang} Steinhardt P J, Wang L and Zlatev I 1999
 \emph{Phys. Rev}. \textbf{D59} 123504
\bibitem{Ng}
Ng S C C, Nunes N J and Rosati F 2001 \emph{Phys. Rev}.
\textbf{D64} 083510
\bibitem{hao3}Hao J G and Li X Z 2003 \emph{Preprint}
astro-ph/0309746, To be published in \emph{Phys. Rev}. \textbf{D}
\bibitem{muk}Armendariz-Picon C, Mukhanov V and Steinhard P J 2001
\textit{Phys. Rev}. \textbf{D63} 103510
\end{thebibliography}
\end{document}